\documentclass{article}

\usepackage{arxiv}

\usepackage[utf8]{inputenc} 
\usepackage[T1]{fontenc}    
\usepackage{lmodern}        
\usepackage{hyperref}       
\usepackage{url}            
\usepackage{booktabs}       
\usepackage{amsfonts}       
\usepackage{nicefrac}       
\usepackage{microtype}      
\usepackage{graphicx}

\title{A systematic machine learning approach to measure and assess biases in mobile phone population data}

\author{
    Carmen Cabrera\textsuperscript{*}
   \\
    Geographic Data Science Lab, Department of Geography and Planning \\
    University of Liverpool \\
  Liverpool, United Kingdom \\
  \texttt{\href{mailto:C.Cabrera@liverpool.ac.uk}{\nolinkurl{C.Cabrera@liverpool.ac.uk}}} \\
   \And
    Francisco Rowe
    \thanks{The authors contributed equally to all the stages of the work.}
   \\
    Geographic Data Science Lab, Department of Geography and Planning \\
    University of Liverpool \\
  Liverpool, United Kingdom \\
  \texttt{\href{mailto:fcorowe@liverpool.ac.uk}{\nolinkurl{fcorowe@liverpool.ac.uk}}} \\
  }

\providecommand{\tightlist}{%
  \setlength{\itemsep}{0pt}\setlength{\parskip}{0pt}}

\usepackage{longtable,booktabs,array}
\usepackage{calc} 
\usepackage{etoolbox}
\makeatletter
\patchcmd\longtable{\par}{\if@noskipsec\mbox{}\fi\par}{}{}
\makeatother
\IfFileExists{footnotehyper.sty}{\usepackage{footnotehyper}}{\usepackage{footnote}}
\makesavenoteenv{longtable}

\NewDocumentCommand\citeproctext{}{}

\makeatletter
 \let\@cite@ofmt\@firstofone
 \def\@biblabel#1{}
 \def\@cite#1#2{{#1\if@tempswa , #2\fi}}
\makeatother
\newlength{\cslhangindent}
\newlength{\csllabelwidth}
\newenvironment{CSLReferences}[2] 
 {\begin{list}{}{%
  \setlength{\itemindent}{0pt}
  \setlength{\leftmargin}{0pt}
  \setlength{\parsep}{0pt}
  \ifodd #1
   \setlength{\leftmargin}{\cslhangindent}
   \setlength{\itemindent}{-1\cslhangindent}
  \fi
  \setlength{\itemsep}{#2\baselineskip}}}
 {\end{list}}
\usepackage{calc}

\usepackage{amsmath}
\DeclareUnicodeCharacter{0301}{\'{}}
\begin{document}
\maketitle

\begin{abstract}
Traditional sources of population data, such as censuses and surveys, are costly, infrequent, and often unavailable in crisis-affected regions. Mobile phone application data offer near--real-time, high-resolution insights into population distribution, but their utility is undermined by unequal access to and use of digital technologies, creating biases that threaten representativeness. Despite growing recognition of these issues, there is still no standard framework to measure and explain such biases, limiting the reliability of digital traces for research and policy. We develop and implement a systematic, replicable framework to quantify coverage bias in aggregated mobile phone application data without requiring individual-level demographic attributes. The approach combines a transparent indicator of population coverage with explainable machine learning to identify contextual drivers of spatial bias. Using four datasets for the United Kingdom benchmarked against the 2021 census, we show that mobile phone data consistently achieve higher population coverage than major national surveys, but substantial biases persist across data sources and subnational areas. Coverage bias is strongly associated with demographic, socioeconomic, and geographic features, often in complex nonlinear ways. Contrary to common assumptions, multi-application datasets do not necessarily reduce bias compared to single-app sources. Our findings establish a foundation for bias assessment standards in mobile phone data, offering practical tools for researchers, statistical agencies, and policymakers to harness these datasets responsibly and equitably.
\end{abstract}


\section{Introduction}\label{introduction}

Traditional data streams, such as the census and surveys have been the
primary official source to provide a comprehensive representation of
national populations in countries worldwide. However, fast-paced
societal changes and emergency disasters, such as climate-induced
hazards and COVID-19 have tested and accentuated weaknesses in
traditional data systems (Green, Pollock, and Rowe 2021). Traditional data systems often
provide data in infrequent and coarse temporal and geographical
resolutions (Rowe 2023). Generally they are expensive to maintain
and operate, and are slow taking months or years since they data are
collected to their release (Rowe 2023). Data collection from
climate- or conflict-impacted areas is generally unfeasible because of
restrictions due to high levels of insecurity and risk
(Iradukunda, Rowe, and Pietrostefani 2025). Yet, fast-paced societal changes require high
frequency, granular and up-to-date information to support real-time
planning, policy and decision making.

At the same time, we have seen the confluence of two diverging trends in
data availability. On the one hand, growing evidence of declining survey
response rates across many countries over the last 20 years is
accumulating (De Heer and De Leeuw 2002; Stedman et al. 2019; Luiten, Hox, and Leeuw 2020). Dwindling
numbers in surveys can represent distorted picture of society
(Luiten, Hox, and Leeuw 2020). On the other hand, significant advances in sensor
technology, computational power, storage and digital network platforms
have unleashed a data revolution producing large trails of digital trace
data (Kitchin 2014). These data are now routinely collected and
stored. They offer spatially granular, frequent and instant information
to capture and understand human activities at unprecedentedly high
resolution and scale, with the potential to produce real-time actionable
intelligence to support decision making (Rowe 2023). Hence,
national statistical offices are actively seeking to integrate these
data into their national data infrastructure (United Nations Statistics Division 2025; UK Research and Innovation 2025).

Mobile phone data (MPD) collected via GPS- and IP-based technology have
become a prominent source of nontraditional data to monitor population
changes. Increasing usage of mobile services on smartphones and wearable
devices have resulted in the generation of large volumes of geospatial
data, offering novel opportunities to advance understanding of spatial
human behaviour, and thus revolutionise research, business and
government decision making and practices (Rowe 2023). MPD are now
a core component of the digital economy, creating new market
opportunities for data intelligence businesses, such as Cuebiq/Spectus,
Safegraph and Locomizer. They have been used to create critical evidence
to support policy making, prominently during the COVID-19 pandemic. In
research, MPD have been used to develop innovative approach to infer
mode of transport (Graells-Garrido et al. 2023), monitor footfall changes
(Ballantyne, Singleton, and Dolega 2022; Hunter et al. 2021), profile daily mobility signatures
(Cabrera-Arnau et al. 2023), sense mobility accessibility
(Graells-Garrido et al. 2021), predict socioeconomic levels (Soto et al. 2011; Blumenstock, Cadamuro, and On 2015), estimate income segregation (Moro et al. 2021), quantify
tourism activity (Raun, Ahas, and Tiru 2016) and estimate migration (Rowe et al. 2024; González-Leonardo et al. 2025) and population displacement (Rowe et al. 2022; Iradukunda, Rowe, and Pietrostefani 2025).

However, the use of MPD present major epistemological, methodological
and ethical challenges (Rowe 2023). A key unresolved challenge is
potential biases in MPD compromising their statistical
representativeness and perpetuate social injustice
(Wesolowski et al. 2013). Biases reflect societal digital and
socioeconomic inequalities. Biases emerge from differences in the access
and use of the MP applications used to collect MPD
(Porter et al. 2012; Wesolowski et al. 2013). Only a fraction of the population
in a geographical area owns a smartphone, and even an smaller share
actively uses a specific MP app. In the UK, for example, 98\%
of the adult population have a MP and 92\% of this population
use a smartphone (Ofcom 2023), but a smaller percentage actively use
Facebook (70\%) or Twitter (23\%) (Statista 2024). Additionally, biases
emerge from differences in the access and use of digital technology
across population subgroups reflecting socioeconomic and demographic
disparities. For instance, wealthy, young and urban populations
generally have greater access and more intensively use of MP
applications, and therefore tend to be over-represented in MPD
(Blumenstock and Eagle 2010; Wesolowski et al. 2013; Schlosser et al. 2021).

The use of biased MPD can thus have major practical and societal
implications. If used uncorrected, MPD reproduce selective patterns of
smartphone ownership and application usage, rendering inaccurate or
distorted representations of human population activity. Such
representations disproportionately reflect behaviours of younger, urban
and higher-income users while underrepresenting marginalised or
less-connected groups (Porter et al. 2012; Wesolowski et al. 2013). Distorted
representations based on biased MPD can thus misguide decision making,
policy and planning interventions, and thus amplify existing
socio-economic disparities. In practice, existing applications of MPD
often use uncorrected population statistics derived from MPD and have
thus been constrained to offer a partial picture for a limited segment
of the overall population. Such data can only afford to provide rough
signals about the spatial distribution of (e.g.~spatial concentration),
trends (e.g.~increasing) and changes (e.g.~low to high) in populations
(Rowe, Neville, and González-Leonardo 2022). Unadjusted, they have cannot provide a full
representation of the overall population.

Efforts have been made to measure and assess biases in aggregate
population counts from digital data sources. Existing analyses typically
measure the extent of bias measuring the system-wide difference in the
representation of population counts from digital platforms and censuses
(Ribeiro, Benevenuto, and Zagheni 2020; Zagheni and Weber 2015; Gil-Clavel and Zagheni 2019). To estimate the
representation of digital data sources, the penetration rate is computed
as the active user base of a digital platform over the census resident
population (Ribeiro, Benevenuto, and Zagheni 2020; Gil-Clavel and Zagheni 2019). Existing analyses
have thus been able to established systematic gender, age and
socio-economic biases in population data obtained via API (or
Application Programming Interface) from social media platforms, such as
Facebook and Twitter/X. However, this approach requires information on
the demographic and socio-economic attributes of the collected sample
and has focused on estimating biases at the country level. Yet, these
attributes are generally unavailable for MPD, and biases may vary widely
across subnational areas. What is missing is an systematic approach to
measure biases in population counts from digital platforms, when
population attributes are unknown, and quantify the geographic
variability in the extent of biases in these data.

To address this gap, this paper aims to establish a standardised
approach to empirically measure the extent of biases in population data
derived from digital platforms, and identify their key underlying
contextual factors across subnational areas. We seek to address the
following research questions:

\begin{itemize}
\tightlist
\item
  What is the relative level of population coverage and bias of MPD
  sources to widely-used traditional surveys?
\item
  To what extent, does the level of population bias vary across
  subnational areas and cluster in particular regions?
\item
  How systematic is the association between larger population biases
  and over-representation of particular population subgroups, such as
  rural, more deprived and elderly populations?
\item
  To what extent, are MP-based population data from multiple
  applications versus single applications associated with lower
  population bias?
\end{itemize}

Our approach proposes a statistical indicator of population coverage to
measure the extent of bias, and uses explainable machine learning to
identify key contextual factors contributing to spatial variations in
the extent of bias. Biases in digital trace data can emerge from
multiple sources, such as algorithmic changes, device duplication and
geographic location accuracy (Rowe et al. 2025). We do not intend to
identify these individual sources of error. We focus on quantifying the
extent of ``cumulative'\,' bias; that is, the resulting bias from the
accumulation of these error sources. We identify this as population
coverage bias. We use data collected from single and multiple MP apps,
and compare their results. As outlined above, we test the extent to
which biases can be mitigated by leveraging information from multiple
apps encompassing a more diverse user population. Specifically, we use
two single-app (i.e.~Facebook and Twitter/X) and two multi-app providers
(i.e.~Locomizer and an undisclosed provider). We focus on the use of
aggregated population counts as this has become a common ethical and
privacy-preserving practice for companies to provide access to highly
sensitive data for social good.

Our study makes two key contributions. Our first contribution is
methodological. We develop and demonstrate a systematic, replicable
approach for assessing the quality of MP--derived population data when
information on population attributes is unavailable. This approach
quantifies population coverage bias at national and subnational scales,
identifies the degree of spatial variation and clustering, and uses
explainable machine learning to identify the contextual drivers of these
biases. By establishing a simple yet robust and transparent indicator of
coverage bias, our approach provides a practical tool to evaluate
digital trace data prior to substantive analysis. In doing so, we
respond directly to recent calls in the human mobility literature for
greater transparency and standardisation in the processing and
validation of digital trace data (Barreras and Watts 2024; Rowe et al. 2025; UK Government 2025). While prior work has
emphasised the fragility of analytical choices and the absence of good
practices in this field, our approach establishes a clear set of
indicators and procedures that can be applied consistently across
sources and geographies. By providing a privacy-preserving and
transparent benchmark, our approach offers a foundation for emerging
standards of good practice in the use of MDP, bridging the gap between
innovative digital signals and the established norms of accountability
that underpin official population statistics.

Our second contribution is substantive. Using our proposed approach, we
present the first systematic cross-platform assessment of biases in MP
app datasets for the UK. We show that coverage bias varies substantially
across sources and geographies, and that demographic, socioeconomic, and
geographic features consistently explain much of this variation. This
evidence provides robust insights into the structural roots of bias,
such as the role of age composition, education, occupation, and rurality
in shaping representation. Importantly, we also demonstrate that MP
multi-app datasets -often assumed to be more representative- do not
inherently guarantee lower bias. Instead, they capture a broader
population base at the expense of introducing more complex and
multidimensional patterns of bias compared with single-app sources.
These findings extend current debates on the quality of human mobility
data by moving beyond descriptive comparisons to provide systematic,
evidence-based assessment. They also directly address concerns raised by
Barreras and Watts (2024) that the lack of validation across datasets
hinders the reliability of policy-relevant metrics. Our results
highlight that improving representativeness requires source-specific
adjustment strategies and cannot rely solely on aggregation across apps,
thereby setting a new standard for empirical evaluation of MP--based
population estimates.

\section{Data}\label{sec-data}

\begin{figure}
\centering
\includegraphics[width=14.5cm,height=7.2cm]{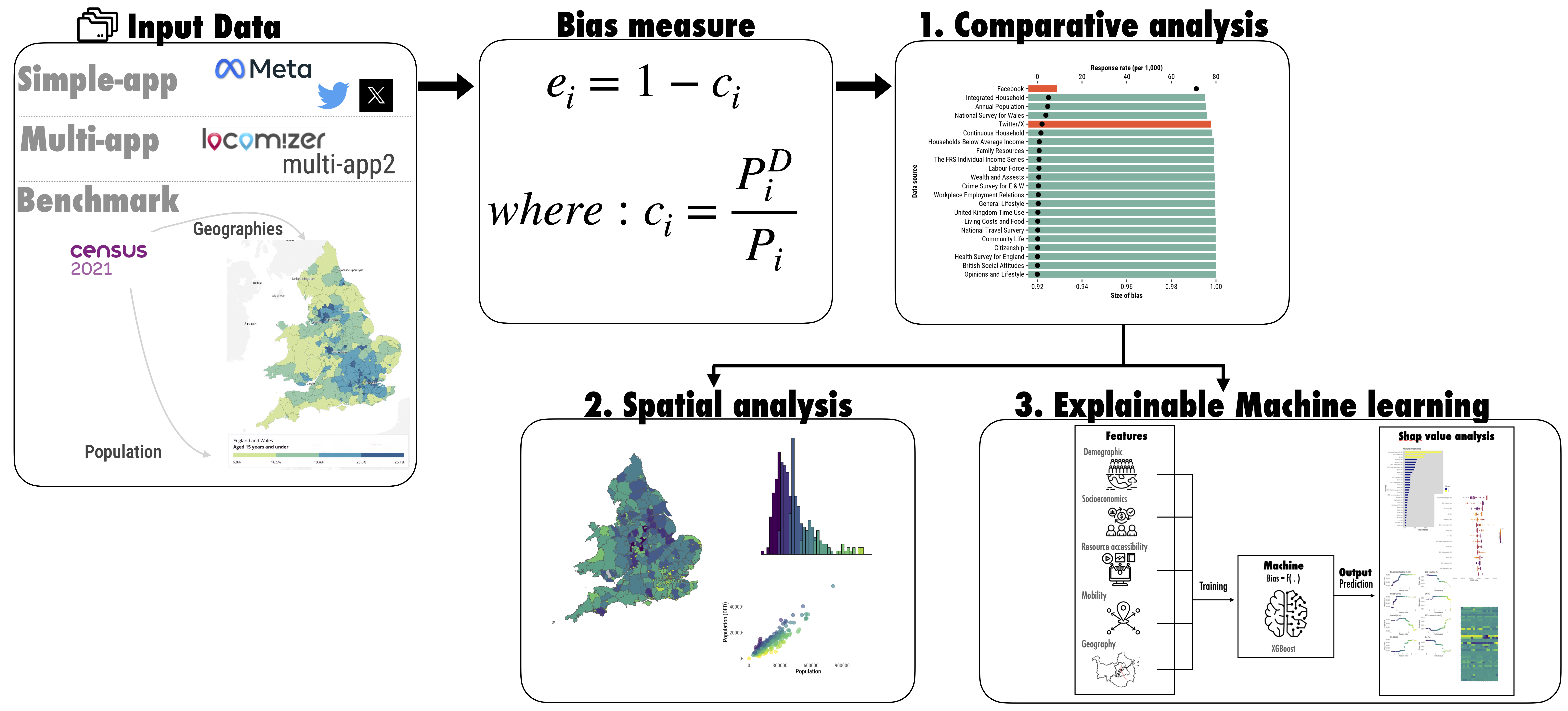}
\caption{Systematic approach to measure and explain biases from MP-based
population estimates.}\label{fig:debias-approach}
\end{figure}

We propose a systematic framework to measure and explain biases in
population count data derived from MPs. Figure
\ref{fig:debias-approach} provides a digramatic representation of our
proposed approach consisting of four stages: (1) measuring coverage
biases; (2) assessing the comparative extent of bias against widely used
survey data; (3) determining the level of spatial variability in bias;
and, (4) identifying key contextual sources of biases across subnational
areas. To illustrate our proposed framework, we draw on four MP datasets
collected in March 2021 to align as closely as possible with the dates
of the most recent census, and thus ensure temporal consistency. As
explained below, the key principle here is that we use census data as
our benchmark, so we assume that full representation is achieved when a
source provides the same level of coverage as the census. We focus on
aggregated population counts, which are commonly used in mobility
research, as a privacy-preserving and ethically responsible data format.
The datasets include sources derived from a single MP application (Meta
and Twitter/X) as well as from multiple MP applications, each capturing
distinct user groups through different data generation mechanisms. These
differences allow us to assess how source characteristics influence
population coverage and representativeness. The multi-application
sources are referred to as Multi-app1, whose provider name cannot be
disclosed due to a non-disclosure agreement, and Multi-app2, provided in
its raw format by the company Locomizer. We conducted analysis at the
Local Authority District (LAD) level given that data from Facebook and
Twitter/X are available at aggregate spatial scales. Table
\ref{tab:data-source} summarises the main characteristics of each
dataset, including the source type, form of data collections, temporal
granularity, temporal coverage, spatial resolution, access method and
data acquisition cost. Further details of access and processing for each
data source are provided next.

\begin{table}[h]
\centering
\includegraphics[width=1\linewidth]{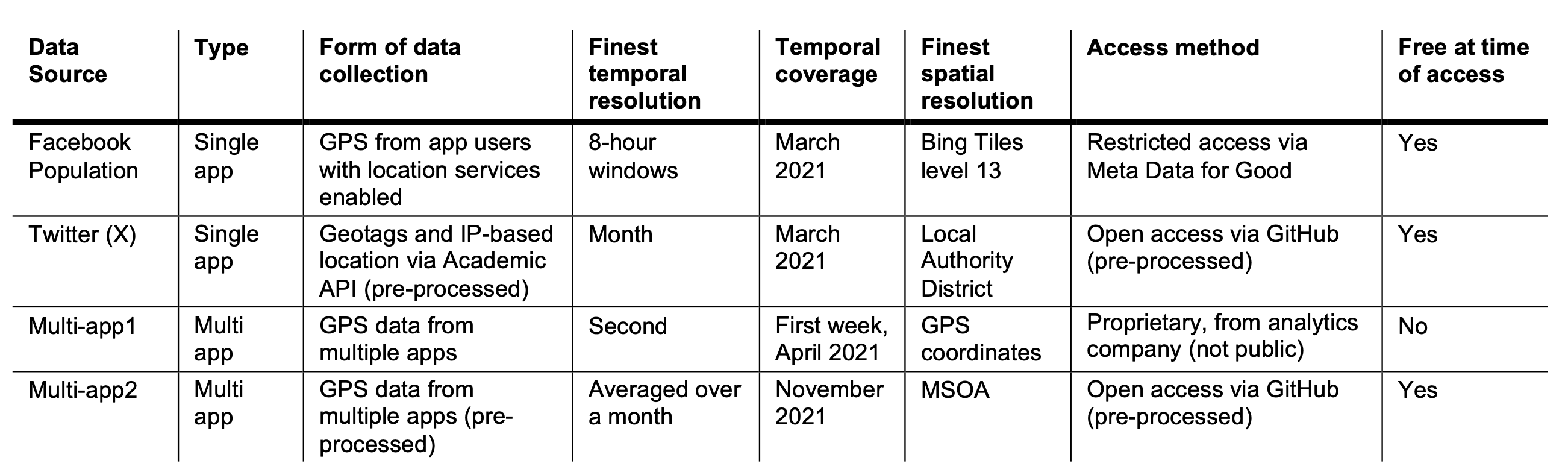}
\caption{Summary description of mobile phone data sources.}
\label{tab:data-source}
\end{table}

\subsection{Meta}\label{meta}

We used the Facebook Population dataset created by Meta and accessed
through their Data for Good Initiative\footnote{\url{https://dataforgood.facebook.com}}. This dataset provides
anonymised aggregate location data from Facebook app accounts, who have
the location services setting activated.
We used data for the UK covering March 2021 when the most recent UK
Census was carried out. Prior to releasing the datasets, Meta ensures
privacy and anonymity by removing personal information and applying
several techniques which include small-count dropping for population
counts under 10, addition of random noise and spatial smoothing using
inverse distance-weighted averaging (Maas 2019).

The dataset includes the number of active Facebook app users, aggregated
into three daily 8-hour time windows (i.e.~00:00-08:00, 08:00-16:00 and
16:00- 00:00). To approximate the resident population, we focus on the
time window corresponding to nighttime hours (00:00--08:00), when users
are more likely to be at home. For the study area, this time window
yields an average of 4.2 million daily user records. Spatially, the
Facebook Population data is aggregated according to the Bing Maps Tile
System (Microsoft 2023). In this study, we used data aggregated
at Bing tile level 13, which corresponds to a spatial resolution of
approximately 4.9 \(\times\) 4.9 km at the Equator (Maas 2019).

To integrate the Facebook and census data, we aggregated the Facebook
Population data by averaging daily population counts to the level of
LADs, to ensure temporal and spatial alignment with official census
data. We provide evidence as Supplementary Material (SM) testing
alternative processing strategies, including averaging over a single
week in March and reversing the order of spatial and temporal
aggregation. These sensitivity provides evidence that our main findings
are robust to variations in the strategy of spatial and temporal
aggregations.

\subsection{Twitter/X}\label{twitterx}

We used an anonymised, analysis-ready dataset of active Twitter/X
accounts in the UK, originally collected via the Twitter Academic API.
We used unique accounts as a proxy for the number of unique users. The
data consists of monthly records for the location of unique Twitter/X
accounts, spatially aggregated across the UK, and is openly available\footnote{\url{https://github.com/c-zhong-ucl-ac-uk/Twitter-Internal-Migration}}.
Geolocation is obtained either directly from geotagged tweets or through
manual geocoding using bounding boxes provided by the API, based on the
IP address of the posting device (for methodological details, see
(Wang et al. 2022)). The full dataset includes approximately 161 million tweets
from February 2019 to December 2021. For this study, we restricted the
analysis to March 2021 to align with the timing of the 2021 UK Census,
during which 125,637 user home locations were identified. Home locations
were assigned to LADs using a frequency-based detection algorithm,
further described in (Wang et al. 2022).

\subsection{Multi-app1}\label{multi-app1}

We sourced data from a location analytics company that collects GPS data
from approximately 26\% of smartphones in the UK. The raw data consist of
anonymised device-level GPS traces collected via a range of smartphone
applications, where users have explicitly granted location-sharing
permissions. We considered the number of devices as a proxy for the
number of unique users, although it could be the case that some users
have more than one device. The dataset spans a 7-day period
corresponding to the first week of April 2021 and includes 443,553,155
GPS records. Although the dataset does not perfectly align with the
official 2021 UK Census date, the temporal proximity ensures a high
degree of comparability.

To infer the place of residence of users, we applied a commonly used
rule-based classification method, following approaches outlined in
(Zhong et al. 2024; Rowe et al. 2025). Specifically, the place of
residence associated with a device is defined as the location with the
highest number of GPS records recorded during nighttime hours (10 PM--6
AM). To be classified as a residence, a location must account for more
than 50\% of the device nighttime records. Furthermore, the number of
nighttime records during the observation period must be at least 2. For
comparability across data sources, all identified residence locations
are aggregated to the level of LADs. Using this method, we detected
1,536,922 home locations.

\subsection{Multi-app2}\label{multi-app2}

Our analysis includes a second source of analysis-ready dataset of
population counts. This dataset is openly-available on GitHub\footnote{\url{https://t.ly/dzlzB}}, and was processed to
identify the home location of users according to the methodology
described in (Zhong et al. 2024). The raw data is collected by a UK-based
data service company, which licenses mobile GPS data from 200 smartphone
apps and applies pre-processing methods to ensure user privacy and
anonymity. The dataset covers the entire UK for November 2021 and
includes inferred home and work locations for 630,946 users. While this
period does not exactly coincide with the 2021 UK Census, the difference
of less than a year was considered sufficiently close for our analysis.
To ensure consistency across datasets, we further process the data by
aggregating it spatially from the Middle Layer Super Output Area (MSOA)
level to the LAD.

\subsection{Census data}\label{census-data}

We used 2021 UK census, aggregated at the LAD level for two main
purposes. First, we used resident population counts derived from the
census as the denominator of the population coverage measure in the
first stage of the methodology (see Methods section). The resident
population count derived from the census was used as the official
benchmark for population counts. The core assumption is that census data provide the ``truth'\,'
resident population count and so deviation from it indicates greater bias. We thus compare population
counts derived from each digital dataset against from derived from the census.
Second, we also draw on a set
of area-based covariates from the census, covering demographic,
socioeconomic, resource accessibility, mobility-related and geographic
characteristics. These variables are detailed in Section \ref{sec-eml}. They are used as predictors in stage three of the
methodology (see Methods section), to investigate and explain the
factors that are most strongly associated with the magnitude and spatial
variation of coverage bias in the digital trace data. This allows us to
analyse representativeness bias in each source of digital data.

\section{Methods}\label{methods}

We introduce a framework to measure and explain biases systematically in
population count data derived from MPs. This framework
consists of three stages. We first introduce a metric to quantify bias
in the form of population coverage within a given geographic area. We
refer to this as ``population coverage bias''. Next, we compute the
population coverage bias metric for different data sources and
subnational geographic areas. We then analyse its variability and assess
its unevenness distribution across geographies. Detecting such patterns
is important for understanding the limits of data applicability, and to
assess whether spatial dependencies should be considered in identifying
the contextual sources of population bias. In a final stage, we analyse
representativeness bias. We do this by using explainable machine
learning to model the variation in coverage bias as a function of key
contextual features derived from the 2021 UK census. This modelling
approach allows us to model the magnitude of bias across areas, and
quantify the relative importance of each covariate, so it is possible to
identify the population characteristics (e.g.~age structure, income
levels, educational attainment) that are most strongly associated with
overrepresentation or underrepresentation in the different sources of MP
app data. As shown above, Figure \ref{fig:debias-approach} provides an overview of the methodological workflow,
which includes data acquisition, bias measurement, comparative analysis
with national surveys, spatial analysis, and bias explanation through
modelling.

\subsubsection{Measuring population coverage bias}\label{measuring-population-coverage-bias}

We defined a metric to quantify the magnitude of coverage bias in each
subnational area. This metric is based on the population coverage of the
dataset, which we compute as the ratio of the population captured by
dataset \(D\) (sample size) in a geographic area \(i\), denoted as \(P_i^D\),
to the total local population of the same area, \(P_i\). Formally, the
coverage \(c_i\) for area \(i\) is given by:

\begin{equation}
c_i = \dfrac{P_i^D}{P_i} \times 100.
\end{equation}

The resulting ratio \(c_i\) is assumed to take values
between \(0\) and \(100\), with 100 representing full population coverage.
If users have multiple accounts, the ratio can exceed \(100\), since the
total sample size could be greater than the local population of area
\(i\).

We then define the size of bias \(e_i\) for area \(i\) as:

\begin{equation} \label{eq:size-bias}
e_i = 100 - c_i
\end{equation}

A value of \(e_i = 0\) indicates a lack of coverage bias, which
corresponds to full population coverage (\(c_i = 100\)). We use this bias
indicator to analyse the magnitude and distribution of coverage bias
across multiple sources of data and geographic areas.

\subsubsection{Identifying spatial patterns of population bias}\label{identifying-spatial-patterns-of-population-bias}

For each data source, we computed the coverage bias metric at the
subnational level and examined its geographic variation. This stage has
two main objectives. The first assesses the extent of geographical unevenness in bias
across subnational areas. The second seeks to determine whether spatial effects are
sufficiently strong to consider them in the subsequent methodological
stage through the inclusion of spatial lag terms in an explainable
machine learning model.

To evaluate the variability of bias across geographies, we first
conducted exploratory analyses using thematic maps and histograms. To
formally test for spatial clustering, we calculate Moran's I statistic
for each dataset. Because Moran's I is sensitive to the definition of
spatial relationships, we evaluate four alternative spatial weighting
schemes: 1) queen neighbourhood, 2) k-nearest neighbours, 3) distance
band, and 4) distance band (Rey, Arribas-Bel, and Wolf 2023).
Comparing results across these schemes enables us to assess the
robustness of clustering patterns to different definitions of the
spatial weight matrix. In the main body of the paper, we report Moran's
I values obtained using scheme 1, as it produces the highest statistic
across datasets when statistically significant, thereby providing the
most conservative test for the presence of spatial clustering. Results
for the other schemes are provided in the Supplementary Information. For
each dataset, we compute the range of Moran's I values across the four
spatial weighting schemes (maximum minus minimum). The largest such
range observed across all datasets is 0.286. This indicates that, while
the exact values of Moran's I vary somewhat with the choice of weighting
scheme, the differences are small and the resulting statistics remain
relatively close.

To assess whether bias is associated with population size, we
examine the relationship between population counts from digital data
sources and census population counts. If bias varies systematically with
population size, we would expect departures from proportionality.
Conversely, an approximately linear relationship through the origin with
a stable slope would suggest that bias is largely independent of
population size. To test this, we generate scatterplots comparing the
two sources and quantify the strength of the relationship using
Pearson's correlation coefficient.

\subsubsection{Identifying sources of representativeness bias via explainable machine learning}\label{sec-eml}

We used explainable machine learning to identify the key predictors of
population bias and how the importance of these predictors varies across
geographical areas. Existing evidence based on social media suggests
that population location data from digital platforms are biased
over-representing urban, wealthy and young-adult populations (Blumenstock and Eagle 2010; Wesolowski et al. 2013; Schlosser et al. 2021).
We therefore modelled our measure of population bias from
Equation \ref{eq:size-bias} as a function of key area-level attributes
reflecting geographical differences in engagement and access to digital
technology across demographic, socioeconomic, resource accessibility,
mobility and geographic factors. Table 2 reports the set of predictors
included in our analysis. We used data from the 2021 census for England
and Wales to measure these predictors.

We used an eXtreme Gradient Boosting (XGBoost) algorithm. XGBoost is an
ensemble that combines outputs from multiple models to produce a single
prediction and represents an efficient and scalable adaptation of the
gradient boosting machine algorithm proposed by (Friedman 2001). It
utilises gradient descent to improve model performance, and decision
trees are built iteratively, with each tree built to minimise the error
residuals of a preceding iteration. XGBoost has been optimised for
scalability and computational efficiency, providing high predictive
accuracy with limited training time (Chen and Guestrin 2016; Nielsen 2016).
XGBoost has also become one of the most widely-used off-the-shelf
machine learning models in applied settings because of its built-in
regularization that mitigates overfitting, sparsity-aware tree
construction and parallelisation efficiency (Chen and Guestrin 2016). It can
accommodate nonlinearities and is robust to multicollinearity
(Chen and Guestrin 2016). We fitted the following XGBoost regression model.

\begin{equation} \label{eq:xgb-model}
\widehat{e}_i 
= \sum_{m=1}^M f_m\bigl(D_i, S_i, H_i, U_i, L_i\bigr),
\quad f_m \in \mathcal{F}
\end{equation}

\(e_i\) is our measure of population bias. \(f_m\) denotes an individual
regression tree from the boosted ensemble \(\mathcal{F}\) and \(M\) is the
total number of trees. The input variables \(D\), \(S\), \(H\), \(U\), \(L\)
represent key demographic, socioeconomic, resource accessibility, mobility and geographic attributes of area \(i\), respectively. Table
\ref{tab:covariates} lists and describes these features. Based on existing literature, Table
\ref{tab:covariates} also describes the expected relationship with population bias. Here it is relevant to highlight that representative subnational data on digital resource accessibility and engagement is not available in the UK. We proxied this via our census-derived measures of resource accessibility. The model iteratively
learns the contribution of each feature to the prediction of the bias
indicator \(e_i\), allowing for complex, nonlinear interactions.

\begin{table}[h]
\centering
\includegraphics[width=1\linewidth]{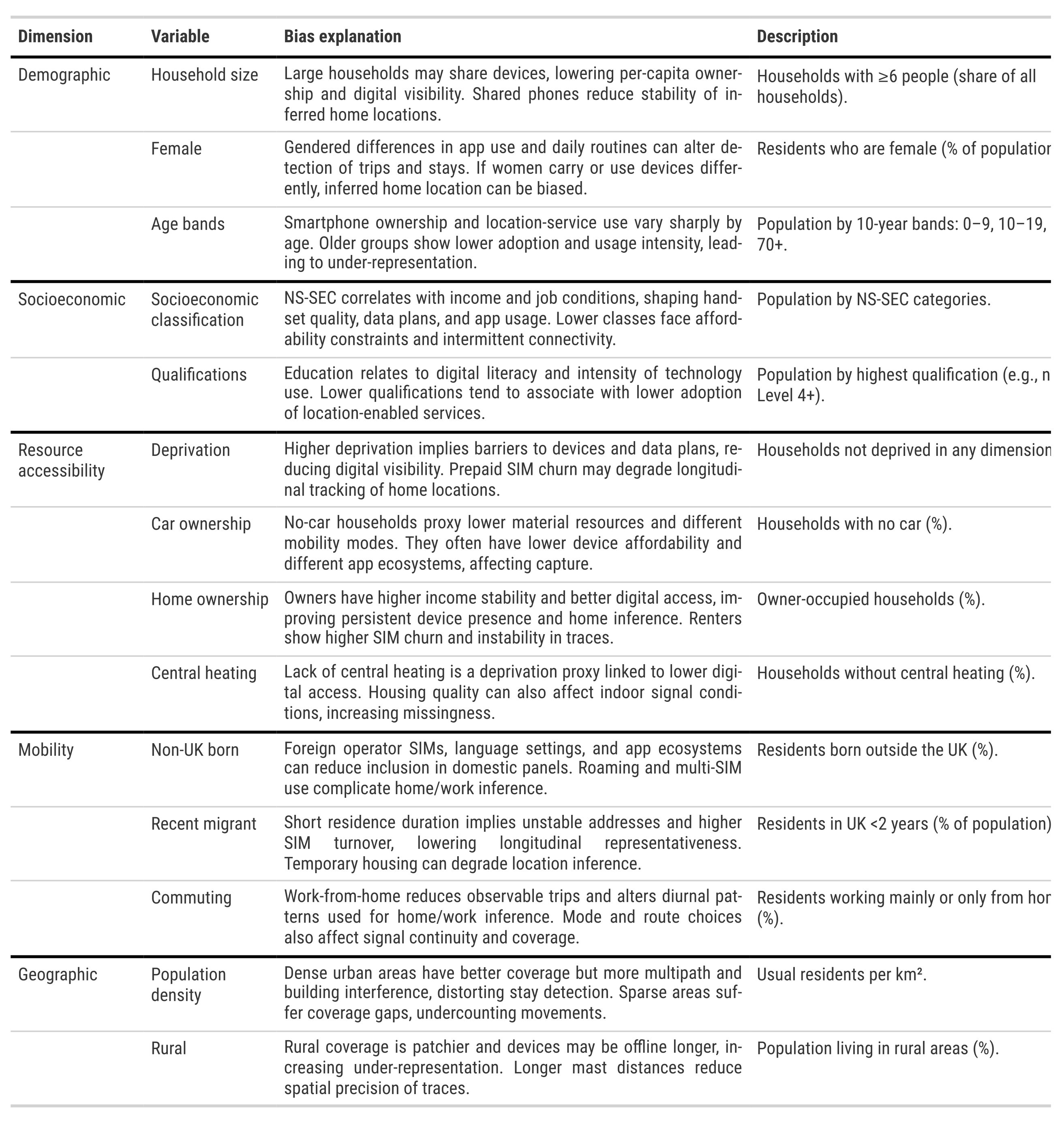}
\caption{Model variable description, expected influence and description.}
\label{tab:covariates}
\end{table}

To implement Equation \ref{eq:xgb-model}, we randomly split the data
into training (80\%) and testing (20\%) sets to ensure robust model
evaluation. We used 10-fold cross validation to train models and
performed grid search over learning rates, tree depths, subsample
ratios, and regularisation penalties to identify optimal
hyperparameters. We applied regularisation penalties including L1
(Lasso) and L2 (Ridge) terms to penalise overly complex trees, promote
feature sparsity, improve model generalisation and mitigate
multicollinearity among predictors. XGBoost's tree-based structure
additionally handles multicollinearity by hierarchically selecting the
most informative splits (Chen and Guestrin 2016). We then fitted a final model on the
full training set using these tuned settings of optimal parameters and
evaluated on the held-out test set. We evaluated models based on the
number of trees minimising the root mean squared error (RMSE), the
convergence of training and test error, and difference between predicted
and observed values.

\section{Results}\label{results}

Next, we illustrate our proposed methodological framework on four
sources of digital data derived from MP apps . As
described in the Data section \ref{sec-data}, these sources include
data for the UK, collected in or around March 2021 to align as closely
as possible with the reference date of the most recent national census.

\subsection{The extent of population coverage bias varies across data sources}\label{the-extent-of-population-coverage-bias-varies-across-data-sources}

\begin{figure}
\centering
\includegraphics[width=14.5cm,height=8.5cm]{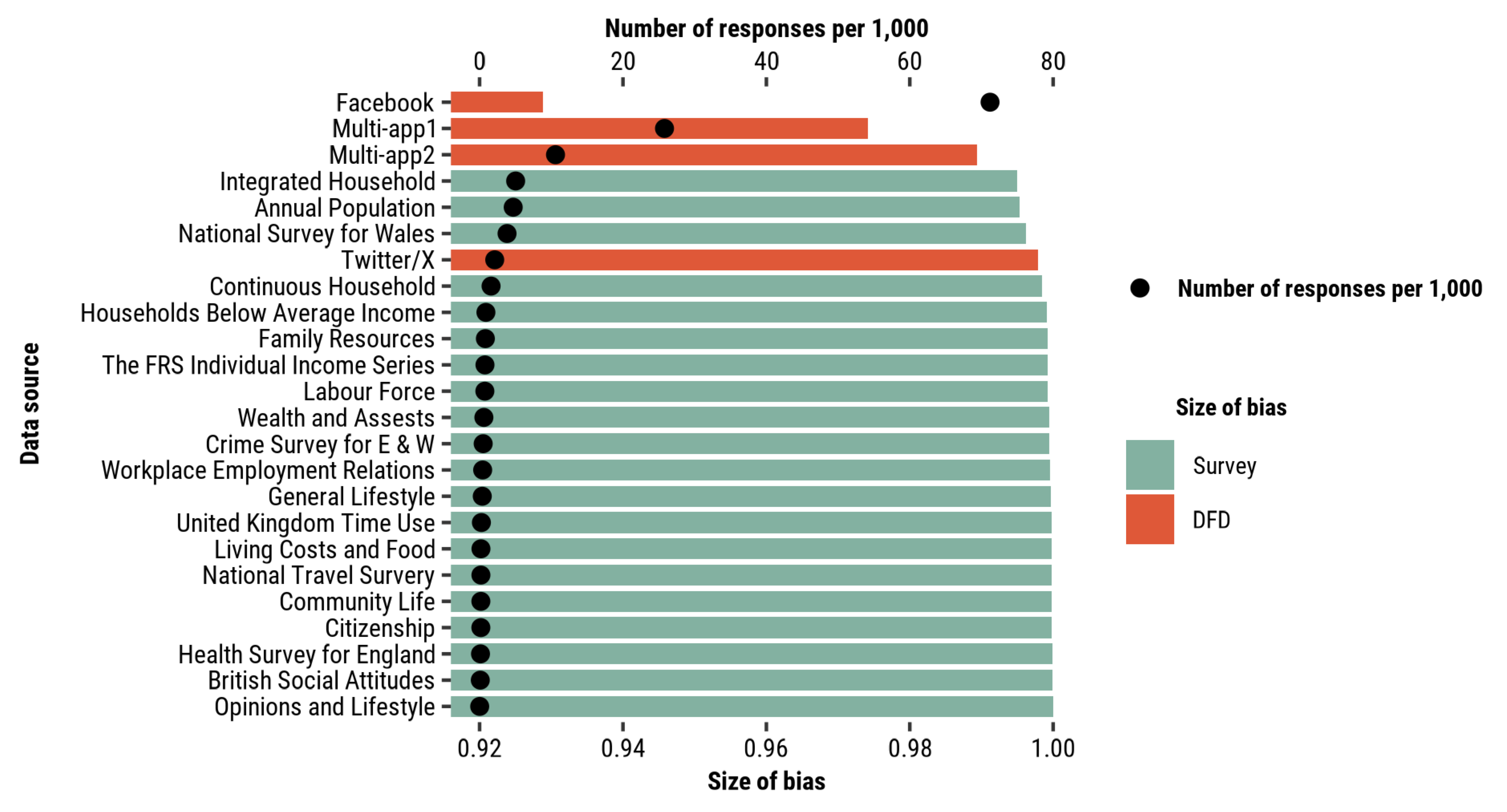}
\caption{Size of coverage bias (bottom x-axis) and population coverage per
1,000 population (top-x-axis) by data
source.}\label{fig:survey}
\end{figure}

Population coverage bias was first computed at the national level for
each MP data source. To contextualise these results, we compared the MP
sources with several widely used traditional datasets, including major
UK surveys available through the UK Data Service
(UK Data Service). Figure \ref{fig:survey} presents these
comparisons across data sources, showing on the top x-axis the
population coverage (expressed as the number of respondents or subjects
per 1,000 people) and on the bottom x-axis the corresponding measure of
coverage bias.

Figure \ref{fig:survey} shows differences between MP data and
traditional survey sources. UK surveys typically achieve coverage of
only a few individuals per 1,000 population, whereas MP datasets can
capture more observations. At the same time, MP data exhibit
comparatively lower national-level coverage bias, suggesting that in
terms of raw size and breadth, digital trace data have strong potential
to support large-scale empirical analyses.

Despite the higher population coverage in MP data, both MP data and
traditional surveys offer complementary strengths and weaknesses.
Traditional surveys are designed to achieve statistical
representativeness, usually, at the national level. Sampling techniques
such as stratified or cluster sampling are applied and, if imbalances
remain, further post-stratification adjustments are implemented
(Lohr 2021). Yet, representativeness can only ever be guaranteed with
respect to a finite set of attributes (e.g., age, gender, income,
region) (Cochran 1977). Moreover, survey data usually provides a
snapshot at one point in time, and offers little visibility into how
coverage and bias vary geographically or dynamically over time.

By contrast, MP data is not collected with sampling strategies and
usually lacks demographic identifiers, which makes it difficult to apply
post-stratification adjustments. It is also generated passively, as a
byproduct of digital interactions, without any guarantee of inclusion
for particular groups. However, the much broader poulation coverage and
fine spatio-temporal resolution of MP datasets provide information that
is usually unavailable through traditional surveys cannot. They allow us
to track how representation varies across regions and through time, and
this is particularly valuable for studying dynamic social processes.

We argue that, even though we do not always have specific demographic
information of the individuals captured through digital trace data, we
can infer some of these characteristics by leveraging the
spatio-temporal granularity of MP data. This is a necessary first step
to understand which population groups might be overrepresented or
underrepresented in different sources of MP data. This information is
necessary to develop subsequent data adjustment strategies that can
improve the representativeness of the data relative to the target
population.

\subsection{Population bias varies widely over space}\label{population-bias-varies-widely-over-space}

To examine coverage bias at subnational levels, we leveraged the
fine-grained geographic resolution of the MP app data sources. Analysing
the distribution of coverage bias across geographies allowed us to
identify uneven patterns of population coverage across areas and assess
whether these patterns exhibit spatial clustering. These assessments
were important both for evaluating the limitations of each dataset for
further research and for deciding whether spatial dependence terms
should be incorporated into the subsequent modelling stage
\ref{sec-explain}.

\begin{figure}
\centering
\includegraphics[width=14.5cm,height=13.5cm]{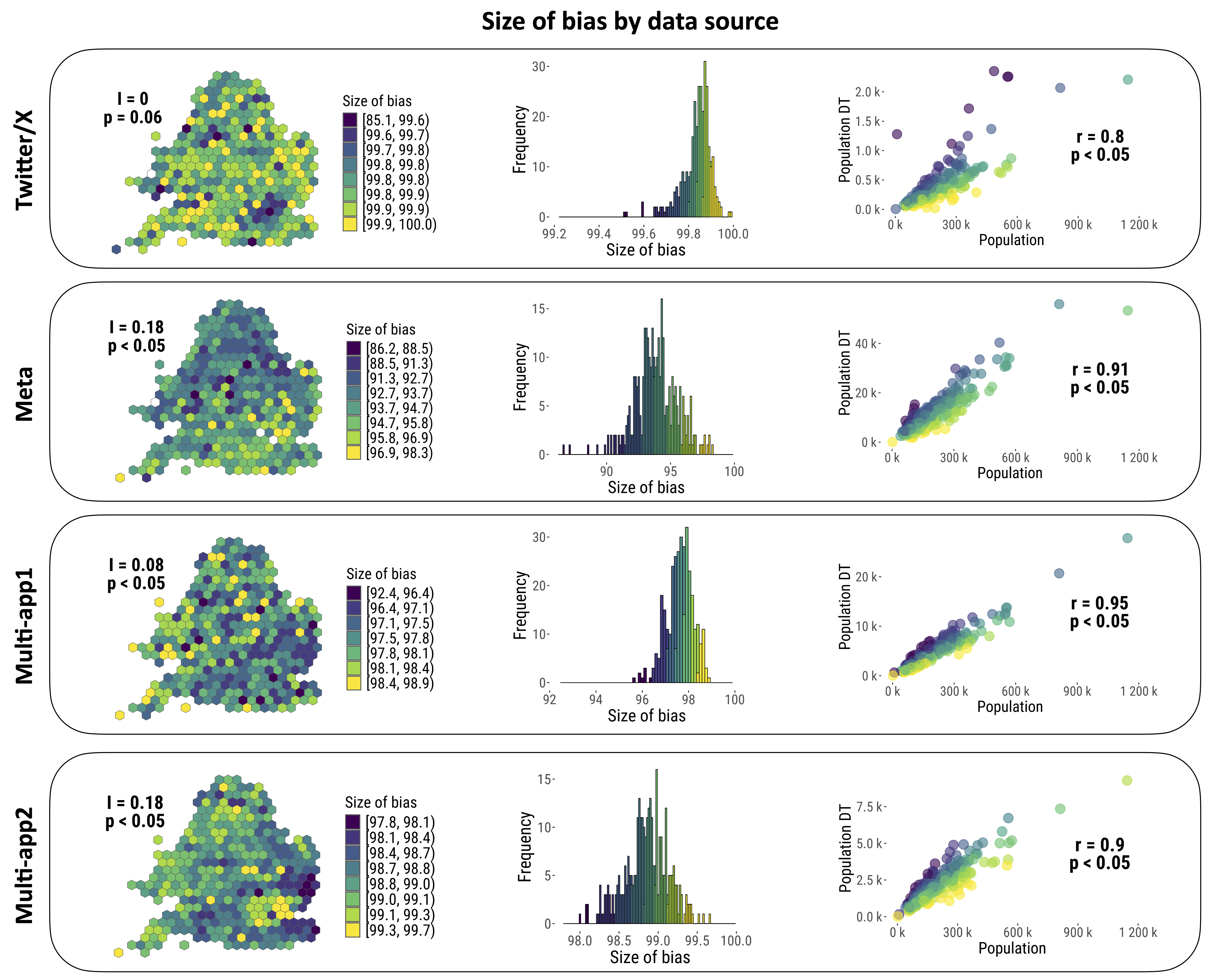}
\caption{Extent and spatial distribution of population bias across local
authority districts in the UK. A. Hexagon map of the size of population
bias across local authority districts. \(I\) represents the Moran's I
and \(p\) denotes the associated p-value. B. Histogram of the
distribution of the size of population bias. C. Scatter plot of the
relationship between population size and size of
bias.}\label{fig:bias-size}
\end{figure}

Figure \ref{fig:bias-size} summarises these analyses for each dataset.
Each row corresponds to a data source and includes three elements. The first shows a
hexagonal cartogram showing the size of coverage bias in each LAD,
alongside the Moran's I statistic and corresponding p-value as a measure
of spatial autocorrelation. The second shows a histogram showing the distribution of
coverage bias across LADs. The third displays a scatter plot comparing population
counts derived from each MP dataset with census population counts for
the same LADs, together with Pearson's correlation coefficient and
p-value.

Our spatial analysis revealed several patterns in the distribution of
coverage bias which are consistent across the MP
datasets. First, all data sources exhibited noticeable geographic
variability in coverage bias, as evidenced by the spread of values in
their respective distributions, as displayed in the histograms. For
example, bias in data from Meta ranges from 86.2 (\(\approx 14\%\)
population coverage) to 98.3 (\(\approx 2\%\) population coverage). This
variability highlights that for a given data source, the degree of
representation can differ substantially between areas.

Second, despite the variability in coverage bias, the maps did not
reveal strong geographic clustering patterns. Bias values fluctuate
across longitude and latitude, and there are no strong north--south or
east--west gradients. This observation was quantitatively supported by
the Moran's I statistics, which are generally statistically significant,
but close to zero. Their small magnitude indicated that spatial
clustering is weak at the LAD scale. Consequently, we concluded that it
is unnecessary to include spatial dependence terms (e.g.~spatial lag) in
the subsequent modelling stage of our framework.

Third, the variability in coverage bias was not explained by the
absolute population size of each LAD. Scatter plots comparing population
counts derived from MP data and from the census revealed strong linear
relationships. This was quantitatively supported by Pearson correlation
coefficients consistently close to one and statistically significant.

In terms of specific patterns of population coverage bias, we observed
some differences across datasets. We found that Twitter/X data shows the
higher values of population coverage bias across MP data sources. The
majority of values of population coverage bias size for this source are
above 99 (\(\approx 1 \%\) population coverage). An exception is the LAD
for City of London, where population coverage bias is 85.1
(\(\approx 15 \%\) population coverage), where a higher number of home
locations was detected. This pattern is likely related to the unique
demographic and spatial context of the City of London. Although it has a
very small resident population, it is a major centre of employment and
commuting. Because home locations are inferred from the most frequently
recorded nighttime position, even a relatively modest number
of residents or temporary visitors can yield an unusually high apparent
population coverage compared to larger LADs. By contrast, data from Meta
displayed the lower values of population coverage bias, ranging from
86.2 (\(\approx 14 \%\) population coverage) to 98.3 (\(\approx 2 \%\)
population coverage). Although Moran's I for Meta data is small
(\(I = 0.18\)), the corresponding map suggested a slight gradient with
LADs in Northern England displaying lower values of bias. Multi-app1 and
Multi-app2 data showed intermediate levels of bias, but with differing
distributions. Notably, data from Multi-app1 shows levels of population
coverage bias as low as 92.4 for some LADs (\(\approx 8 \%\) population
coverage), contrasting with the lowest levels of population coverage
bias for Multi-app2, 97.8 (\(\approx 2\%\) population coverage).
Multi-app1 and Multip-app 2 also show different spatial patterns in the
distribution of population coverage bias across LADs.

Taken together these findings suggested that there is spatial
variability in coverage bias at the LAD level, but this variability is
not explained by physical geographic location or population size.
Instead, the observed patterns are more likely explained by other
area-level characteristics, such as demographic or socioeconomic
composition. We examined these factors in the third stage of our
analysis.

\subsection{Key contextual features explain population biases}\label{sec-explain}

\begin{figure}
\centering
\includegraphics[width=14.5cm,height=16cm]{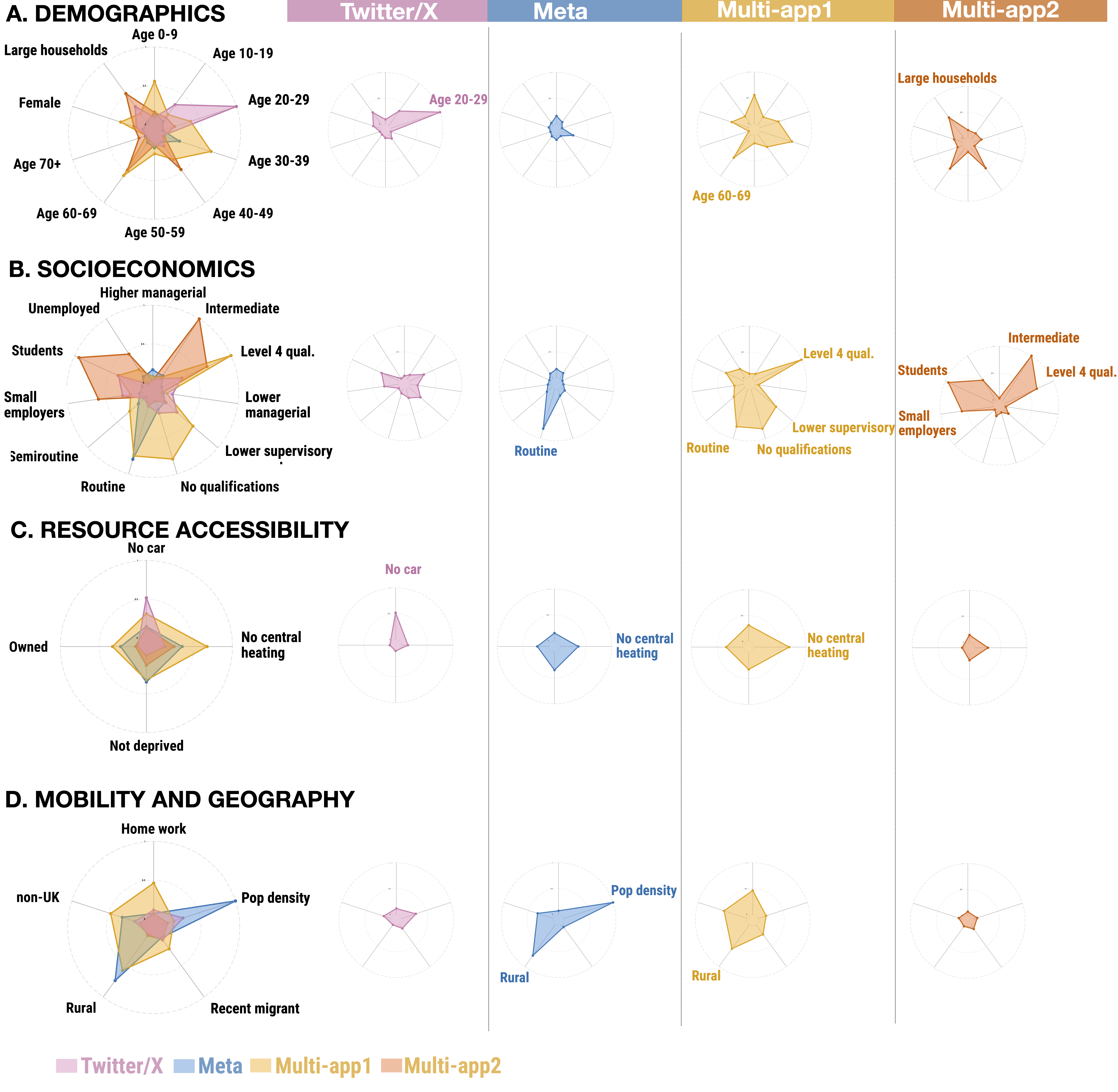}
\caption{\textbf{Radial charts illustrating the importance of, and contribution to,
explaining differences in population biases across local authority
districts.} The importance is estimated based on SHAP feature
importance scores, calculated as the average absolute SHAP value per
feature using an XGBoost machine learning algorithm. Average absolute
SHAP values were normalised across individual data using minimum and
maximum scores, to ensure comparability across data sources. First
column displays the estimates for all data sources and model features.
Subsequent columns highlight features scoring SHAP values over 0.5
within a 0-1 range for individual data
sources.}\label{fig:radialplots}
\end{figure}

We assessed the contribution of key contextual features to explaining
spatial variations in population biases across demographic,
socioeconomic, resource accessibility, mobility and geographic domains.
Figure \ref{fig:radialplots} displays the relative importance of each
individual model feature, representing the average absolute SHapley
Additive exPlanations (SHAP) value per feature including in our XGBoost
model described in Section \ref{sec-eml} \textbf{{[}REF{]}}. SHAP values are
standardised by data source based on minimum and maximum scores to
enable comparability. Figure \ref{fig:radialplots} highlights features
with standardised importance scores over 0.5 for individual data source.

The results reveal wide variability in the key predictors of population
biases across digital data sources. Demographic and socioeconomic
features appear as the most important predictors of population bias in
Twitter/X data. Two features standout -the shares of population aged
20-29 and no car ownership- reflecting the over-representation of the
young adult populations or population with access to resources on active
Twitter/X users. In addition to demographic and socioeconomic features,
resource accessibility and geographic attributes also display some of
the largest contributions to explain population biases across Meta,
Multi-app1 and Multi-app2 platforms. Coupled to the share of population
in routine occupations and lacking central heating, geographical factors
particularly population density and rurality report the highest SHAP
averages, contributing to explain the spatial variability of
Meta-derived population biases. Socioeconomic features standout as the
most important predictors of population biases across both multi-app
platforms, though differences exist. The population share with high or
no qualification, and working in low skill occupations emerge as the
most important features in explaining population biases in Multi-app1.
This is in addition to the population share aged 60-69, residing in
households lacking central heating and living in rural areas. For
Multi-app2-derived estimates, the share of student population, working
in intermediate-level jobs, small employers, having a Level 4
qualification and living in large households score the highest average
SHAP values. The variability in feature importance ranking across data
sources reflects differences in the contextual features that contribute
to explaining biases in their respective population estimates.

\begin{figure}
\centering
\includegraphics[width=14.5cm,height=12.5cm]{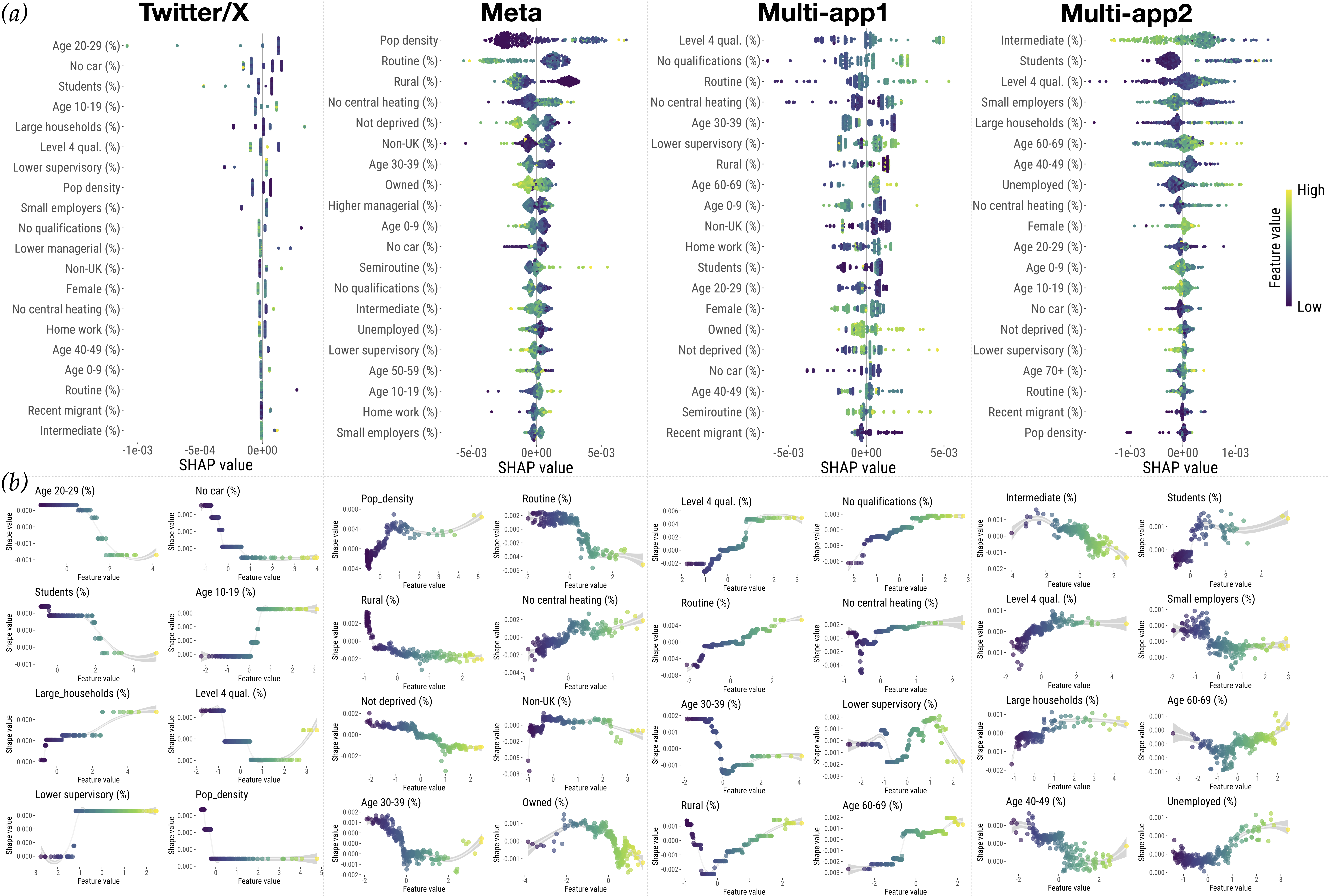}
\caption{\textbf{Top ranked model features contributing to explaining population
biases across local authority districts from an XGboost model.} (a)
beeswarm plots of SHAP feature values displaying the relative importance
and direction of influence of the top 20 contextual features on the
extent of population bias. Features are ranked by their mean absolute
SHAP value, with colours indicating feature values (low to high). (b)
SHAP dependence plots for the top six features based on their mean
absolute SHAP value, illustrating the marginal effect of variation in
each predictor on population bias. Local polynomial regression modelling
was used to represent local trends with 95\% confidence intervals. Each
point represents a local authority
district.}\label{fig:shap-plots}
\end{figure}

Expanding this evidence, Figure \ref{fig:shap-plots} depicts the way
these contextual features contribute to increasing or reducing the
extent of population bias across LADs. Figure \ref{fig:shap-plots}.a
shows the top 20 features based on their average SHAP value from the
highest to the lowest. Figure \ref{fig:shap-plots}.b displays top six of
these features revealing how changes in feature values contribute to
changes in population bias, with colour encoding feature values. For
instance, the first column of plots show that areas with larger shares
of population aged 20-29 and lacking central heating tend to have lower
population bias compared to those with larger shares in
Twitter/X-derived estimates. In contrast, areas with larger shares of
population aged 10-19 are associated with higher Twitter/X-based
population bias, potentially reflecting a limited number of active
Twitter/X users in this age range. Of the features highlighted in Figure
\ref{fig:radialplots}, Figure \ref{fig:shap-plots} indicates that
population estimates derived from Meta tend to have larger biases in
areas with higher population density and greater percentages of people
lacking central heating. Operating in the opposite direction, larger
shares of people working in routine-level jobs and living in rural areas
display lower Meta-based population bias, reflecting greater engagement
with Facebook among these communities. For Multi-app1-derived estimates,
biases are larger for areas with greater shares of population with Level
4 or no qualification, working in routine-level jobs, lacking central
heating, living in rural communities and population aged 60-69. For
Multi-app2-derived estimates, biases are greater for areas with smaller
shares of people working in intermediate-level jobs and self-employed in
small businesses, but displaying larger percentages of students, people
with Level 4 qualification and living in large households.

Figure \ref{fig:shap-plots} also reveals systematic complex nonlinear
shapes in the association between contextual features and population
bias. We identified three types of nonlinear relationships. First is
curvilinear associations in the way of U-shape or inverse U shapes.
These involve patterns of population biases decreasing at low values of
contextual features, increasing at medium values and reducing again - or
the reverse. A distinctive pattern of these associations is their
curvature representing a reversal in the direction of the relationship
between population bias and a contextual feature. Meta-based population
density and Multi-app2-based intermediate feature estimates represent
prominent curvilinear relationships. A second pattern takes the form of
S-shaped associations. The distinctive feature of these associations is
their multiple phase composition displaying a different pattern of
population biases at low and high end values of the contextual feature
relative to middle range feature values. Twitter/X-based 20-29 age
estimates, for example, display high and unchanging population bias at
low feature values, highly variable bias at mid values and low but
increasing population bias at high values. A third pattern is
threshold/stepwise associations representing sharp changes at a cut-off.
The distinctive feature of this pattern is sharp ``steps'' or thresholds
where the relationship between population bias and a contextual feature
changes abruptly. Population biases would remain flat at low values and
then jump and plateau at higher values. Our Multi-app1 estimates for
Level 4 qualification represent a clear illustrative of this type of
association displaying flat small population bias at low feature values
but, then jump and remain high at higher feature values. These shapes do
not appear to be data source specific and vary in unpredictable ways
across variables.

\section{Discussion}\label{discussion}

MPD have become a key data asset to understanding population dynamics in
near real-time at high spatial and temporal resolution. Yet, biases have
remained a major barrier eroding their trust and wider adoption. A key
limitation to tackle this issue has been the lack of a standardised
method to estimate and analyse coverage biases in MPD-derived population
data. In this study, we sought to contribute a systematic and
generalisable approach to measure and analyse coverage biases in
MP-derived population data, when access to individual-level demographic
attributes are lacking. Implementing this approach on UK data, we
presented evidence showing that MDP tend to provide greater population
coverage than widely used traditional surveys. Our findings also
revealed that coverage biases vary markedly across data sources and
subnational areas, that multi-app data sources does not necessarily
offer broader coverage than single-app sources, and that demographic,
socioeconomic and geographic features consistently explain a large share
of spatial variation in coverage bias. Additionally, we identified
distinctive nonlinear effects in the association between contextual
features and coverage bias, suggesting that such biases cannot be
addressed with simple linear adjustment.

These findings carry wide-ranging implications for the use of MPD in
research and official statistics. First, the framework that we propose
offers a replicable approach for evaluating the quality of MP-derived
population counts when information on population characteristics is
unavailable, establishing a baseline for transparent quality assessment.
A key contribution is that it allows researchers and statistical
agencies to move beyond anecdotal assumptions about representativeness
and instead adopt a systematic, quantitative measure of bias that can be
applied consistently across data sources and geographies. In practice,
this means that MP datasets can be assessed before they are used in
substantive analyses, enabling the identification of contexts where the
data are reliable and those where caution is warranted. By being
source-agnostic and privacy-preserving, the framework provides a common
standard of evaluation that can be integrated into official workflows
for producing population estimates, mobility indicators or crisis
response statistics. In doing so, it helps bridge the gap between
innovative but often opaque digital trace data and the established norms
of transparency and accountability that underpin official statistics
(Rowe et al. 2025).

Second, the evidence provided here underscores the need for
source-specific adjustment strategies to mitigate data biases.
Single-app datasets, such as those derived from social media platforms,
tend to display biases that are largely concentrated around a narrower
set of characteristics, most notably age profiles, patterns of digital
engagement, and certain geographic features. These biases, while still
consequential, are comparatively easier to model and adjust for through
established weighting or calibration techniques. Our evidence goes some
way in validating the use of existing simple re-weighting scheme to
mitigate biases based on age and sex structures. By contrast, multi-app
datasets, which aggregate signals from a diverse set of platforms,
capture broader segments of the population but at the cost of
introducing more complex and multidimensional biases. These include
systematic under- or over-representation linked to age and geography,
and also to education, occupation, income and household structure. As a
result, correcting multi-app data requires adjustment strategies that go
beyond conventional post-stratification, demanding more sophisticated
statistical tools that can simultaneously account for multiple
intersecting dimensions of inequality. Such tools may include multilevel
reweighting schemes, Bayesian hierarchical modelling (Rampazzo et al. 2021) ,
data-fusion (Graells-Garrido et al. 2023) or difference-in-difference
approaches (Zagheni and Weber 2012) that integrate auxiliary information from
surveys or administrative records. Such distinction suggests that
efforts to improve the representativeness of MP-derived statistics
should be tailored to the nature of the data source, recognising that
multi-app datasets, while promising in their breadth. Multi-app datasets
pose greater methodological challenges and cannot be assumed to be
inherently less biased than single-app data.

Third, our study reinforces the view that digital trace data should not
be treated as neutral reflections of population activity but rather as
selective windows shaped by persistent digital divides. Digital
engagement is deeply patterned by age, gender, socioeconomic status and
geography, meaning that the traces people leave behind are
systematically unequal. Treating these data as representative without
adjustment risks reproducing and amplifying existing inequalities in the
evidence base used for public policy. By contrast, explicitly
recognising the partial and selective nature of MPD allows researchers
to design strategies to correct for these divides, improving both
validity and fairness. This is particularly important in applications
with direct policy relevance, such as crisis response, public health
monitoring, transport planning and infrastructure investment, where
decisions informed by biased data could inadvertently exclude or
disadvantage already vulnerable groups. Ensuring that MPD are used
critically and responsibly therefore requires integrating bias
assessments and corrections into routine analytical practice, thereby
maximising their potential to contribute to equitable decision-making.

Despite these contributions, our study has limitations that should be
acknowledged. First, although we benchmarked against the 2021 UK census,
not all countries have access to such high-quality and up-to-date
reference data. In many regions, particularly in Africa censuses are
infrequent, often collected only every five to ten years, and sometimes
of variable quality (Dindi and Stiegler 2025). To enhance the transferability of our
framework, future work could draw on global gridded population datasets
such as WorldPop (Tatem 2017), which provide continuous, spatially
explicit population estimates that may serve as a useful alternative
benchmark, and a large set of covariates to inform spatial analysis of
bias source identification. Second, while our analysis is based on
aggregated counts, this does not necessarily restrict the assessment of
demographic-specific biases. MPD rarely contain direct information on
individuals' characteristics, meaning that traditional demographic
benchmarking is not feasible based on consistent information from the
original source. Our approach therefore represents a key advantage, as
it provides a systematic way to interrogate and inform adjustment
methods for bias precisely in the absence of demographic attributes.
Third, the reliance on area-based covariates introduces the risk of
ecological fallacy, as heterogeneity within local units may be obscured.
This highlights the need for more spatially granular analyses where data
availability permits. Such analysis is possible if access to more
granular spatial MPD is available. Finally, we examined four datasets.
Further research is required to assess whether the patterns we document
hold across additional providers, different time periods and diverse
national contexts.

Future research should build on our proposed approach in at least three
directions. First, applications in low- and middle-income countries,
where mobile penetration is uneven, are crucial to assess how bias
manifests in settings of greater digital inequality. Second, comparative
work across datasets and providers is needed to identify which sources
are most representative under different conditions and to establish
standards of validation and robustness. Third, integrating our approach
with emerging adjustment methods --- such as data fusion with household
surveys, imputation techniques and post-stratification weighting ---
offers a pathway to correct biases and improve representativeness.
Advances in trusted research environments and synthetic data generation
may also provide new avenues to interrogate raw data more transparently
while safeguarding privacy. Taken together, these efforts will be vital
to ensure that MPD can be harnessed as a timely, granular and adaptive
complement to censuses and surveys, enhancing the capacity of
researchers and policymakers to monitor population distributions in an
era of accelerating social and environmental change.

\section{Conclusion}\label{conclusion}

This paper has introduced a systematic approach to quantify and explain
coverage and representativeness biases in mobile phone application data
when information on population attributes is absent. Applying the
framework to four datasets for the UK, we presented evidence that MPD
can achieve higher overall coverage than traditional surveys, yet
substantial biases persist across sources and subnational areas. We
showed that these biases are systematically associated with demographic,
socioeconomic and geographic characteristics, often in complex nonlinear
ways, and that multi-app datasets, while broader in coverage, present
more complex patterns of bias than single-app data. By establishing a
transparent and replicable approach, our study provides a foundation for
assessing the quality of digital trace data and offers guidance for
designing adjustment strategies to improve their representativeness. In
doing so, we take a step towards enabling MPD to complement censuses and
surveys as reliable sources for research and policy. Realising this
potential will require continued validation across diverse contexts, the
integration of bias-adjustment techniques, and collaborative efforts
between data providers, researchers and statistical agencies to ensure
that the societal benefits of MPD are equitably realised.

\section*{References}\label{references}
\addcontentsline{toc}{section}{References}

\phantomsection\label{refs}

\bibliographystyle{unsrt}
\bibliography{arxiv-preprint.bib}

\end{document}